\documentclass[twocolumn, twocolappendix]{aastex63}

\usepackage{graphicx}

\submitjournal{ApJ}

\shorttitle{Observations of the Current Sheet Heating in X-ray during a Solar Flare}
\shortauthors{Reva et al.}
\begin{document}

\title{Observations of the Current Sheet Heating in X-ray during a Solar Flare}

\correspondingauthor{Anton Reva}
\email{reva.antoine@gmail.com}

\author[0000-0003-4805-1424]{A.A Reva}
\affiliation{Space Research Institute, \\
Profsoyuznaya 84/32, \\
117997, Moscow, Russia}

\author[0000-0002-5448-8959]{S.A. Bogachev}
\affiliation{Space Research Institute, \\
Profsoyuznaya 84/32, \\
117997, Moscow, Russia}

\author[0000-0002-9601-1294]{I.P. Loboda}
\affiliation{Space Research Institute, \\
Profsoyuznaya 84/32, \\
117997, Moscow, Russia}

\author{A.S. Ulyanov}
\affiliation{Space Research Institute, \\
Profsoyuznaya 84/32, \\
117997, Moscow, Russia}

\author[0000-0002-7303-4098]{A.S. Kirichenko}
\affiliation{Space Research Institute, \\
Profsoyuznaya 84/32, \\
117997, Moscow, Russia}

\begin{abstract}
In the solar corona, magnetic reconnection occurs due to the finite resistivity of the plasma. At the same time, resistivity leads to ohmic heating. Therefore, the reconnecting current sheet should heat the surrounding plasma. This paper presents experimental evidence of such plasma heating caused by magnetic reconnection. We observed the effect during a C1.4 solar flare on 16 February 2003 at the active region NOAA 10278, near the solar limb. Thanks to such a location, we successfully identified all the principal elements of the flare: the flare arcade, the fluxrope, and, most importantly, the presumed position of the current sheet. By analyzing the monochromatic X-ray images of the Sun obtained by the \textit{CORONAS-F}/SPIRIT instrument in the \ion{Mg}{12} 8.42~\AA\ spectral line, we detected a high-temperature ($T \geq$~4~MK) emission at the predicted location of the current sheet. The high-temperature emission appeared during the CME impulsive acceleration phase. We believe that this additionally confirms that the plasma heating around the current sheet and magnetic reconnection inside the current sheet are strongly connected.
\end{abstract}


\section{Introduction} \label{sec:intro}

Plasma in the solar corona has such a low resistivity that its motion can be treated in the approximation of the ideal magneto-hydrodynamic (MHD). In this approximation, the connectivity of the magnetic field is conserved: two points that belonged to the same field line will continue to belong to the same field line during plasma motion \citep[see chapter 2 in][]{Priest2014}.

Nonetheless, plasma in the solar corona has non-zero resistivity, and, therefore, the connectivity of the magnetic field lines can change. The most important manifestation of this process is magnetic reconnection: a mutual annihilation of two magnetic lines of opposite polarities at the `magnetic separator.' 

Changes of the coronal magnetic field induce the electric current inside the separator. This current prevents reconnection of the magnetic field lines. If the magnetic field continues to change, the separator will bifurcate into a current sheet: a thin layer of electric current. Eventually, due to the finite resistivity of plasma, the induced current will slowly dissipate, and the magnetic structure will slowly relax to a potential configuration.  This process is called `steady reconnection' \citep[see chapter 1 in][]{Somov2006}.

If the magnetic energy is accumulated faster than it is dissipated by ohmic heating, the current sheet size will increase. Eventually, the current sheet can reach such a size that it becomes unstable, and the whole amount of accumulated energy will be released catastrophically through the `impulsive reconnection.'

\begin{figure}[t]
\centering
\includegraphics[width = 0.47\textwidth]{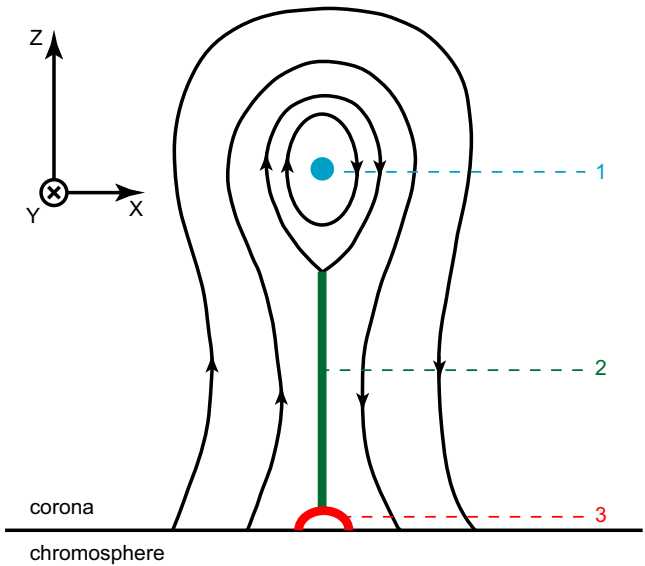}
\caption{Standard flare/CME model. 1) the fluxrope; 2) the current sheet; 3) the flare arcade.}
\label{F:standard_model}
\end{figure}

Impulsive reconnection is a central element of the standard model of a solar flare \citep{Carmichael1964, Sturrock1966, Hirayama1974, Kopp1976}. In this model, before the flare begins, the active region has the following configuration: a loop arcade, a fluxrope (cylindrical twisted magnetic structure) above the arcade, and a current sheet between the arcade and the fluxrope (see Figure~\ref{F:standard_model}). The flare starts when the current sheet becomes unstable, and impulsive reconnection begins inside it. The reconnection process produces two plasma flows, one of which pushes the fluxrope up and may lead to a coronal mass ejection (CME). At the same time, inside the reconnection region, the electrons are accelerated by the induced electric field. The accelerated electrons move along the magnetic field lines towards the chromosphere, where they slow down, heat plasma, and produce a bremsstrahlung hard X-ray (HXR) emission. The heated plasma fills the magnetic loops above the chromosphere, making them visible in soft X-ray and EUV spectral ranges. 

The current sheet is a vital element of the standard flare model. At the same time, it is one of the hardest objects on the Sun to observe. One of the main reasons is that there are no effective ways to directly measure magnetic fields or electric currents in the solar corona. Another reason is that the current sheet is a very thin structure. It has a huge size in the $Y$ and $Z$ directions in Figure~\ref{F:standard_model}, but negligible size along the $X$ axis (several meters, or even several cm). So the plasma emission inside the current sheet is expected to be negligible compared to the emission of the surrounding coronal plasma.

Despite this, several authors reported signatures of the current sheet observed during solar flares. In the imaging observations, a long thin linear structure above the flaring active region is usually interpreted as a current sheet signature \citep{Lin2005, Savage2010, Reeves2011, Zhu2016, Seaton2017}. Besides that, an elongated double Y-shaped dark structure can appear below the CME core in the Fe~171~\AA\ images \citep{Reva2016b}. In the spectroscopic observations, researchers consider a high-temperature emission from the presumed location of the current sheet---below the CME core and above the flaring active region---as a signature of the current sheet \citep{Ciaravella2002, Ko2003, Ciaravella2008}. Recently, \citet{Warren2018} presented multi-wavelength observations of the current sheet using both imaging \citep[AIA;][]{Lemen2012} and spectroscopic  \citep[EIS;][]{Culhane2007} instruments.

In all the examples listed above, the authors reported, not the direct observations of the current sheet, but rather its indirect observational signatures. Among such indirect evidence, one of the most important is the plasma heating in the vicinity of the current sheet caused by the ohmic heating due to the finite resistivity of the coronal plasma.

Mechanisms of such heating are actively studied with the theoretical models and the numerical MHD simulations. These studies showed that ohmic heating \citep{Reeves2010, Reeves2019}, adiabatic compression \citep{Birn2009, Reeves2019}, and turbulence \citep{Ye2020} can effectively heat plasma inside the current sheet.  Some of the thermal energy accumulated inside the current sheet may leak away due to thermal conduction, which will heat the surrounding plasma and widen the visible size of the current sheet \citep[`thermal halo';][]{Yokoyama1998, Seaton2009, Reeves2010}. Using MHD simulations, \citet{Reeves2010} showed that thermal conduction could leak up to 50~\% of the energy released during the reconnection. 

\begin{figure*}[t]
\centering
\includegraphics[width = \textwidth]{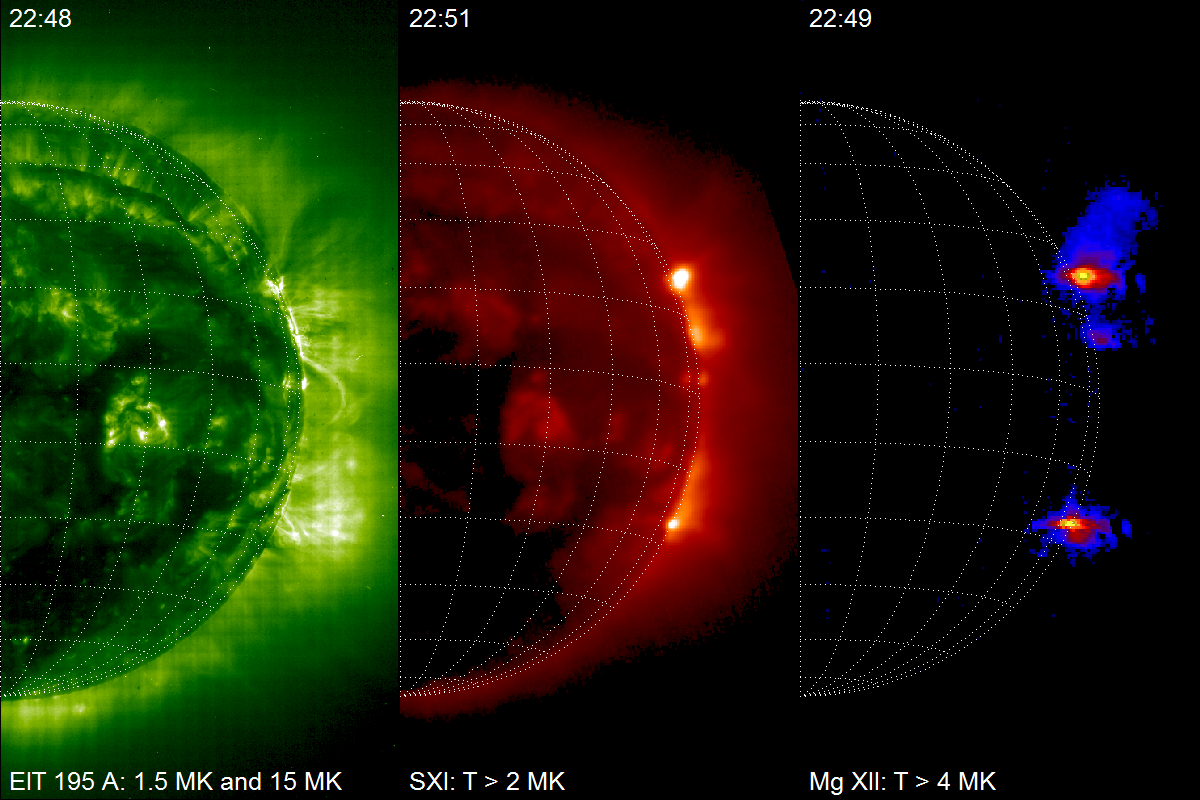}
\caption{Comparison of the images obtained by the EIT 195~\AA\ (left), SXI (middle), and the \ion{Mg}{12} spectroheliograph (right). The images were taken on 16 February 2003.}
\label{F:mg_eit_sxi}
\end{figure*}

For many reasons, it is hard to experimentally study the relationship between the current sheet heating and magnetic reconnection rate. First of all, as mentioned above, the observations of current sheets (even indirect) are rare. Secondly, most solar telescopes---such as AIA/\textit{SDO} or XRT/\textit{Hinode} \citep{Golub2007}---cannot detect high-temperature plasma in a monochromatic mode. The hot plasma emission in their images is mixed with a low-temperature background, which complicates the analysis of the heating. 

In this work, we report reliable signatures of the plasma heating up to temperatures higher than 4~MK detected in the vicinity of the coronal current sheet in a monochromatic mode. We also experimentally confirmed the relationship between the plasma heating and the reconnection rate, which we derived from the observation of CME acceleration. In section~\ref{sec:data}, we describe the experimental data used in the research. In section~\ref{sec:results}, we present the obtained results; then, in section~\ref{sec:discussion}, we discuss them and make the conclusions.

\begin{figure*}[t]
\centering
\includegraphics[width = \textwidth]{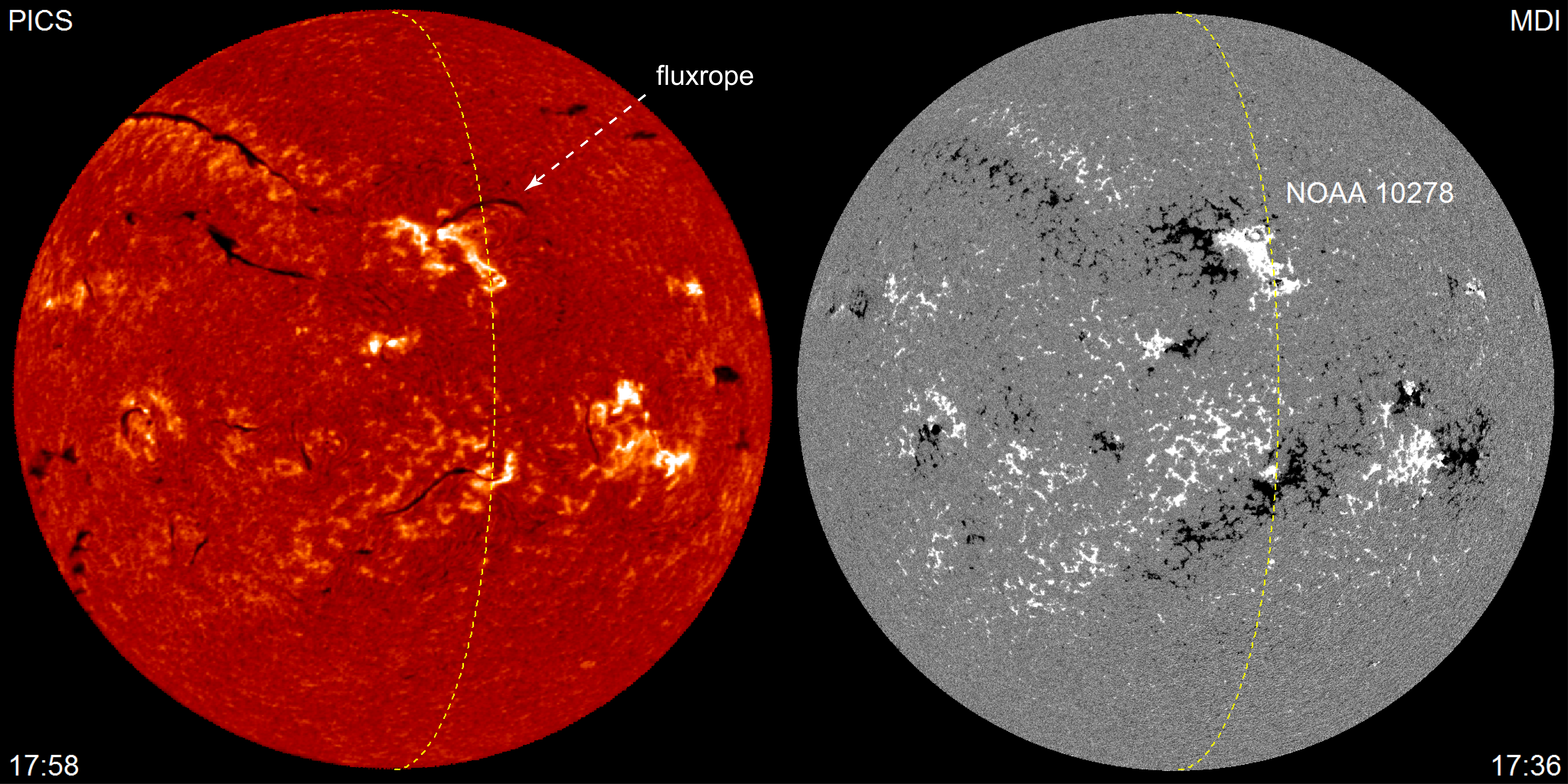}
\caption{Active region NOAA 10278 on the disc. Left: H$\alpha$ image obtained with the PICS telescope. Right: magnetogram obtained with MDI. Images were taken on February 11, 2003. The yellow dashed line indicates the position of the limb at 23:00~UT on February 16, 2003.}
\label{F:halpha_mdi}
\end{figure*}

\section{Experimental Data} \label{sec:data}

\begin{table*}[bht]
\centering
\caption{List of instruments.}
\begin{tabular}{lll}
\hline
\hline
                                                  & Instrument                     &  Reference                  \\
\hline
Location and dynamics of high-temperature plasma  & \ion{Mg}{12} spectroheliograph & \citet{zhi03a}              \\
                                                  & SXI                            & \citet{Hill2005, Pizzo2005} \\ 
Location of the HXR sources                       & RHESSI                         & \citet{Lin2002}             \\
CME structure and dynamics                        & EIT                            & \citet{del95}               \\
                                                  & TRACE                          & \citet{Handy1999}           \\
                                                  & Mk4                            & \citet{Elmore2003}          \\
                                                  & LASCO                          & \citet{Brueckner1995}       \\
Magnetic field configuration                      & MDI                            & \citet{Scherrer1995}        \\
Prominences and filaments in H$\alpha$            & PICS                           & DOI: 10.5065/D65719TR       \\  
\hline
\end{tabular}
\label{T:Instruments}
\end{table*}

The event (flare and CME) that we have studied in this paper occurred on 16 February 2003 near the western edge of the Sun. To study it, we used several instruments to get most of the information about plasma heating, details of the reconnection process, and CME structure. The full list of data used in the study is presented in Table~\ref{T:Instruments}.

The most important instrument for our study was the \ion{Mg}{12} spectroheliograph that operated on board the \textit{CORONAS-F}/SPIRIT satellite from 2001 till 2003 \citep{ora02, Zhitnik2002}. Considering that the instrument may not be well known to some readers, we will briefly describe some of its important features.

The instrument is an imaging spectroheliograph based on Bragg-crystal optics. It obtained monochromatic X-ray images of the solar corona in the \ion{Mg}{12} 8.42~\AA\ spectral line. During the selected period of observations, the spectroheliograph worked with a 2 min cadence and a binned resolution of 8$^{\prime\prime}$. 

The main feature that distinguishes the \ion{Mg}{12} spectroheliograph from other imaging instruments is its temperature selectivity. The \ion{Mg}{12} 8.42~\AA\ line produces a noticeable signal only at temperatures higher than 4~MK. So, the corresponding images clearly outline the high-temperature plasma on the Sun without any contribution from the low-temperature background (see Figure~\ref{F:mg_eit_sxi}). This gives an effective way to study the plasma heating processes in many objects: large-scale flares \citep{Grechnev2006, Urnov07, Reva2015}, CMEs \citep{Kirichenko2013, Reva2017}, and even microflares \citep{Reva2012, Kirichenko2017a, Kirichenko2017b, Reva2018}.

It is important to note that the instrument is equipped with a crystal mirror and, therefore, had dispersion. In the \ion{Mg}{12} images, each point of the Sun looks like a short profile of the \ion{Mg}{12} 8.42~\AA\ spectral line. To lessen this effect, we numerically deconvolved the \ion{Mg}{12} images. For more details, see Appendix~\ref{A:prep}.

\begin{figure*}[thb]
\centering
\includegraphics[width = \textwidth]{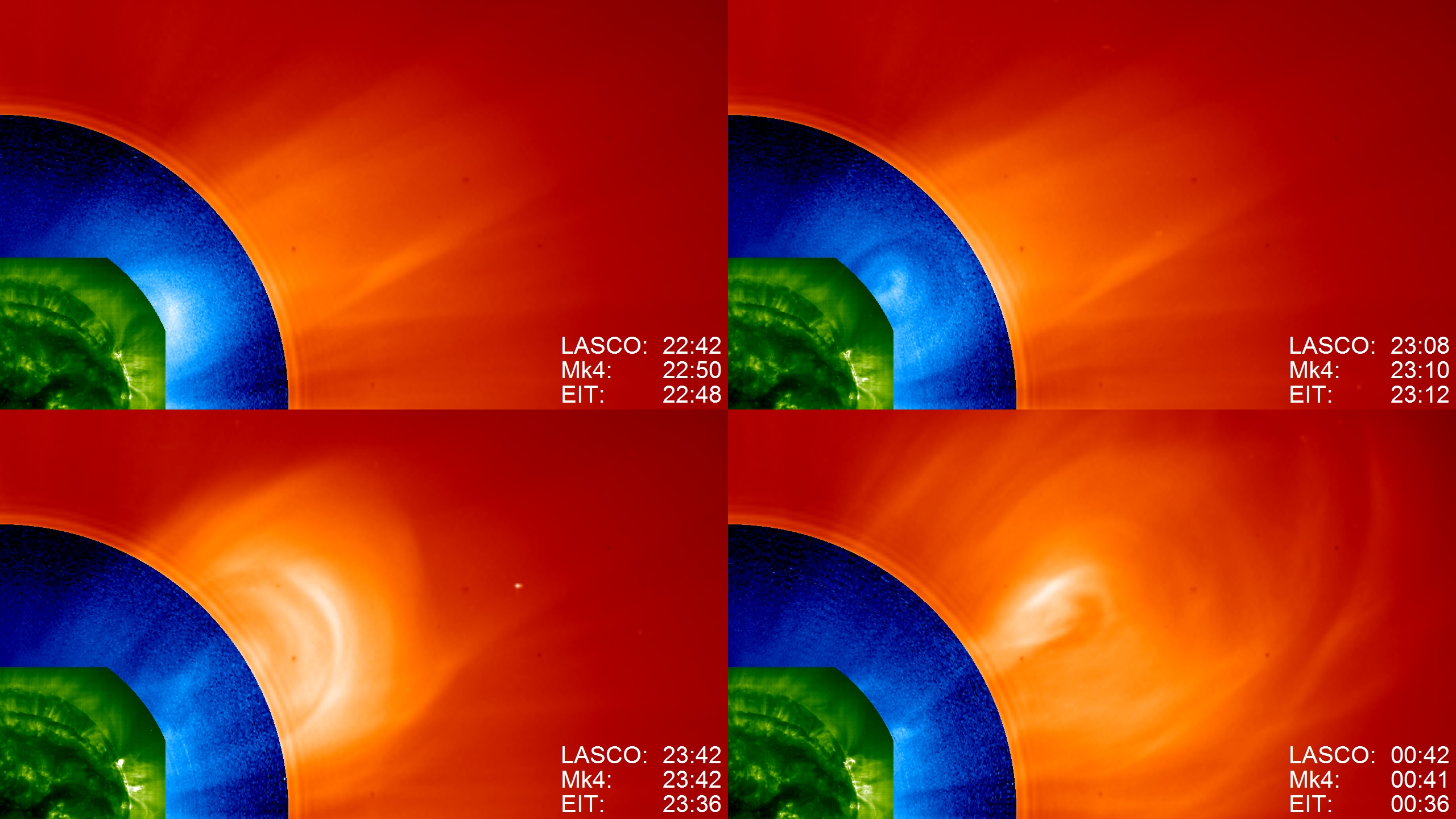}
\caption{Evolution of the CME. Green: EIT telescope; blue: Mk4 coronagraph; red: LASCO C2 coronagraph.}
\label{F:matreshka}
\end{figure*}

The pointing system of the \textit{CORONAS-F} spacecraft had a significant residual jitter. To correct it, we used data from the Solar X-ray Imager \citep[SXI;][]{Hill2005, Pizzo2005} that worked on board the \textit{GOES-12} satellite. The telescope provided full-disk soft X-ray solar images in the 6-60~\AA\ wavelength range with a spatial resolution of $\approx$~10$^{\prime\prime}$ and a 5$^{\prime\prime}$ pixel size. The `Be-thin' channel of SXI is sensitive to the same temperatures as the \ion{Mg}{12} spectroheliograph, but with a noticeable contribution of low-temperature background. The orientation of the Sun in the SXI images is known. Using cross-correlation, we determined the shift between the \ion{Mg}{12} images and the `Be-thin' SXI images. Then we shifted the \ion{Mg}{12} images  by the corresponding value to correct the jitter. 

During the selected period of observations, all of the SPIRIT telemetry was allocated to the \ion{Mg}{12} data. This improved the cadence of the observations, but, as a result, the data of other instruments of the SPIRIT complex were not available.

Below we briefly describe other instruments used in this research.

The {\it Reuven Ramaty High Energy Solar Spectroscopic Imager} \citep[RHESSI;][]{Lin2002} observes HXR spectra from 3~keV to 17~MeV. Using Fourier-based methods, RHESSI can synthesize HXR images in the same spectral range.

The Extreme ultraviolet Imaging Telescope \citep[EIT;][]{del95} on the \textit{Solar and Heliospheric Observatory} \citep[\textit{SoHO};][]{Domingo1995} takes solar images at the wavelengths centered at 171, 195, 285, and 304~\AA. The instrument has a pixel size of $2.6^{\prime\prime}$ and a spatial resolution of $5^{\prime\prime}$. The EIT had two observational modes: synoptic and 'CME watch.' In a synoptic mode, it takes images in all four channels every 6 hours. In the `CME watch' mode, the telescope takes images in the 195~\AA\ channel every 12~min. 

The Transition Region And Coronal Explorer \citep[TRACE;][]{Handy1999} is a space-based telescope that observes the Sun in EUV and white-light. It had a limited field of view ($8.5^{\prime} \times 8.5^{\prime}$) but a high spatial resolution (1$^{\prime\prime}$).

Mk4 coronameter is a ground-based instrument located at Mauna Loa Solar Observatory \citep[DOI: 10.5065/D66972C9; ][]{Elmore2003}. It builds images of the solar corona from 1.14 to 2.86~$R_\odot$ in the white light (700--900~nm) with a spatial resolution of 5.95$^{\prime\prime}$ and a cadence of 3 min. 

Large Angle Spectroscopic Coronagraph \citep[LASCO;][]{Brueckner1995} is a set of white-light coronagraphs that observe solar corona from 1.1~$R_\odot$ up to 30~$R_\odot$ (C1, 1.1--3~$R_\odot$; C2, 2--6~$R_\odot$; C3, 4--30~$R_\odot$). In 1998, LASCO C1 stopped working, and for this research, only C2 and C3 data are available. Coronagraph C2 has a resolution of $11^{\prime\prime}$, and C3 has a resolution of $56^{\prime\prime}$.

The Michelson Doppler Imager \citep[MDI;][]{Scherrer1995} on the \textit{SOHO} satellite maps the line of sight component of the photospheric magnetic field with a  $4^{\prime\prime}$ resolution and 90-min cadence. 

Polarimeter for Inner Coronal Studies (PICS, DOI: 10.5065/D65719TR) is a ground-based instrument located at Mauna Loa Solar Observatory. It takes H$\alpha$ images (6563~\AA) with a field of view of 2.3~$R_\odot$, a spatial resolution of 2.9$^{\prime\prime}$, and a cadence of 3 minutes.

\begin{figure*}[t]
\centering
\includegraphics[width = \textwidth]{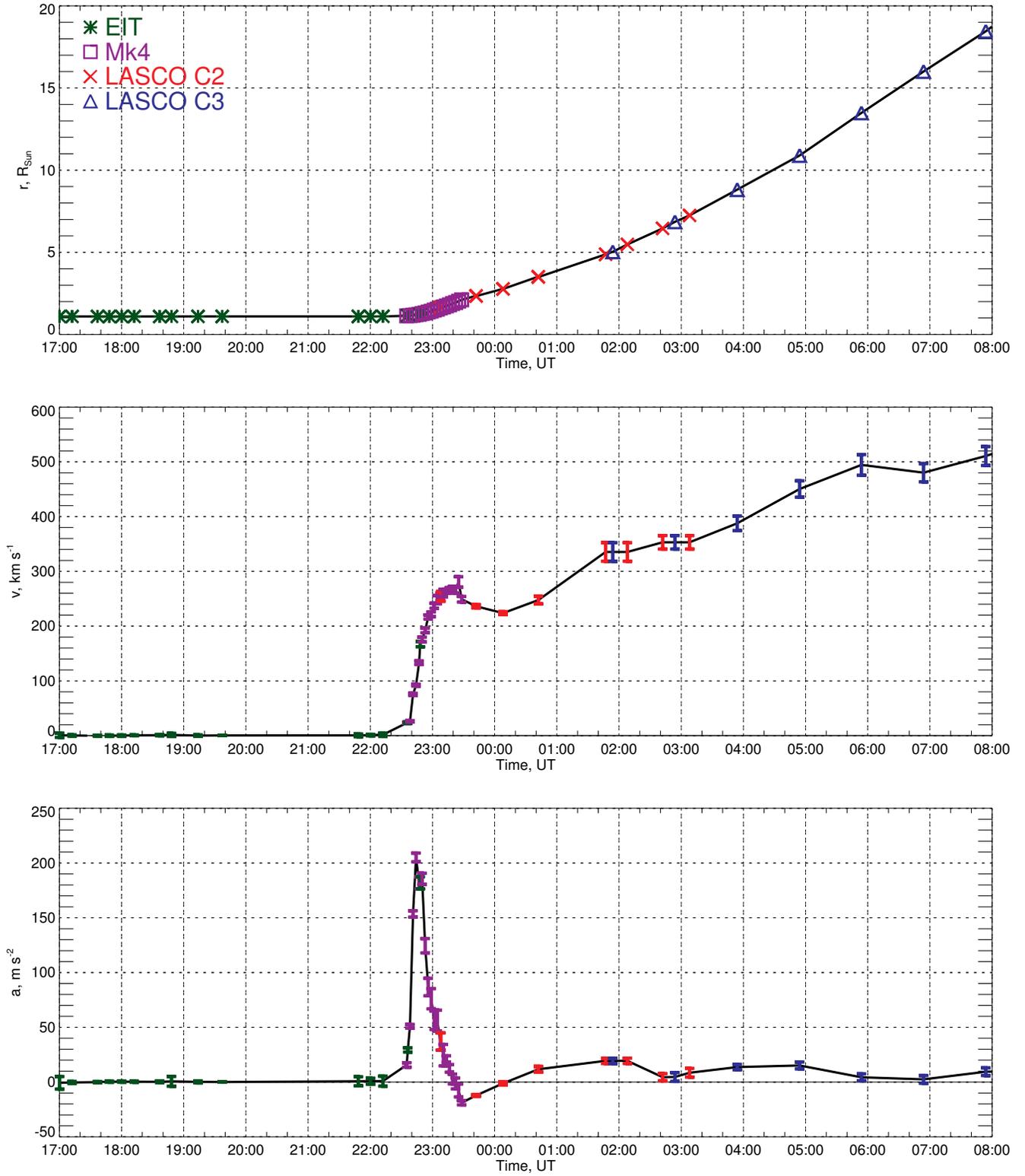}
\caption{Kinematics of the CME core. Top: the distance between the CME core and Sun's center; middle: the CME core's velocity; bottom: the CME core's acceleration. Green: data of the EIT telescope; purple: data of the Mk4 coronagraph; red: data of the LASCO/C2 coronagraph; blue: data of the LASCO/C3 coronagraph.}
\label{F:kinematics}
\end{figure*}

\begin{figure*}[t]
\centering
\includegraphics[width = \textwidth]{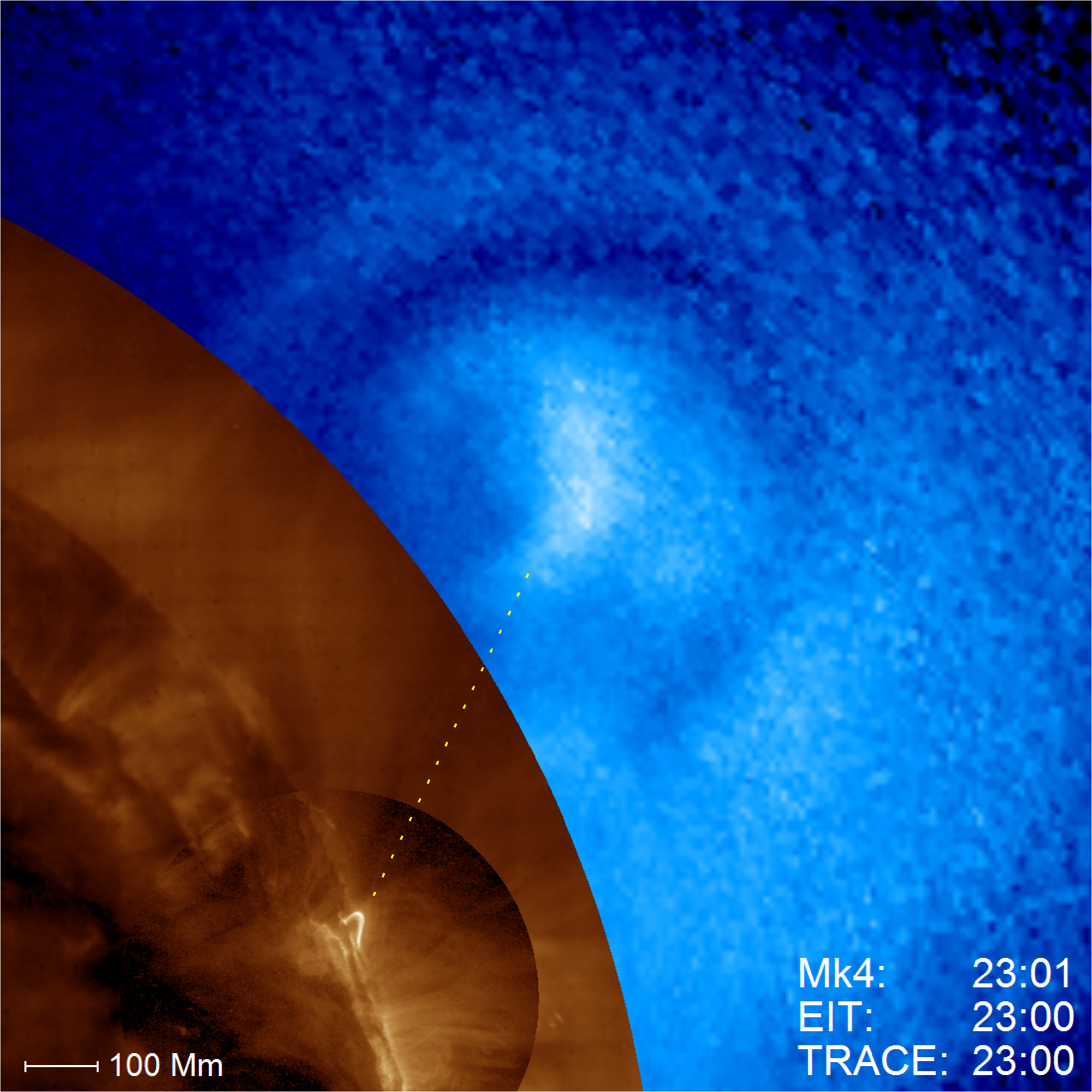}
\caption{CME topology. Copper inner circle: TRACE image; copper outer circle: EIT image; blue: Mk4 image. The dotted line marks the presumed location of the current sheet. The images were taken on 16 February 2003.}
\label{F:trace_eit_mk4}
\end{figure*}

\section{Results} \label{sec:results}

\subsection{Flare Topology and Dynamics}

The studied event was a typical solar flare associated with a CME, which developed in full agreement with the standard views on how a solar flare should evolve. This is very important for our study because it is thanks to the standard configuration of the flare we are sure about where the current sheet was located.

Taking this into account, let us describe the flare topology and dynamics. The pre-flare configuration (5 days before the flare) is shown in Figure~\ref{F:halpha_mdi}, where the left panel is the H$\alpha$ image, and the right panel is the corresponding MDI magnetogram. The flare occurred in the active region NOAA 10278. On 11 February 2003, this active region was approximately in the center of the solar disk. The most prominent feature of the active region seen in H$\alpha$ images (left panel) was a filament that was slightly tilted to the East-West direction and was located between two areas of opposite polarities. We match this filament to the fluxrope of the further CME.

This filament was seen in H$\alpha$ images all five days from February 11 until February 16, when the active region reached the solar limb. At this moment, the region took the same position as in our sketch for the standard flare/CME model (see Figure~\ref{F:standard_model}). So, this was the best projection to observe the current sheet.

Approximately at this time, the CME started to erupt. To study this process, we used synthetic images combined from EIT data obtained in EUV and two white light images: one from the Mk4 coronameter and the second one from the LASCO C2 coronagraph (see Figure~\ref{F:matreshka}). The CME had a classic 3-part structure in white-light images: bright core, dark cavity, and bright frontal loop \citep{Illing1985, Webb1987}. As we said above, we identify the CME core with the filament (fluxrope) that was seen in the H$\alpha$ images before the CME \citep{Illing1986}. In this case, the fluxrope should be aligned along the $Y$ axis in Figure~\ref{F:standard_model}.

Observation of the CME motion gives an indirect way to measure the reconnection rate during a solar flare. Generally, we can expect a simple relationship: the faster is the reconnection rate, the faster is the CME acceleration. 

Taking this into account, we measured the CME coordinates during its motion (for details of the measurement method, see Appendix~\ref{A:kinematics}). The result---the CME height, velocity, and acceleration as a function of time---is shown in Figure~\ref{F:kinematics}. From these plots, we see that CME evolution consists of three main phases:

\begin{enumerate}
\item before $\approx$~22:35, the CME structure was stable;
\item from $\approx$~22:35 to $\approx$~23:10, the CME impulsively accelerated from 0 to 250~km~s$^{-1}$; the acceleration peaked at $\approx$~200~m~s$^{-2}$;
\item after $\approx$~23:10, the CME gradually accelerated to $\approx$~500~km~s$^{-1}$ during 9 hours with an acceleration of $\approx$~10--20~m~s$^{-2}$.
\end{enumerate}

\begin{figure*}[t]
\centering
\includegraphics[width = \textwidth]{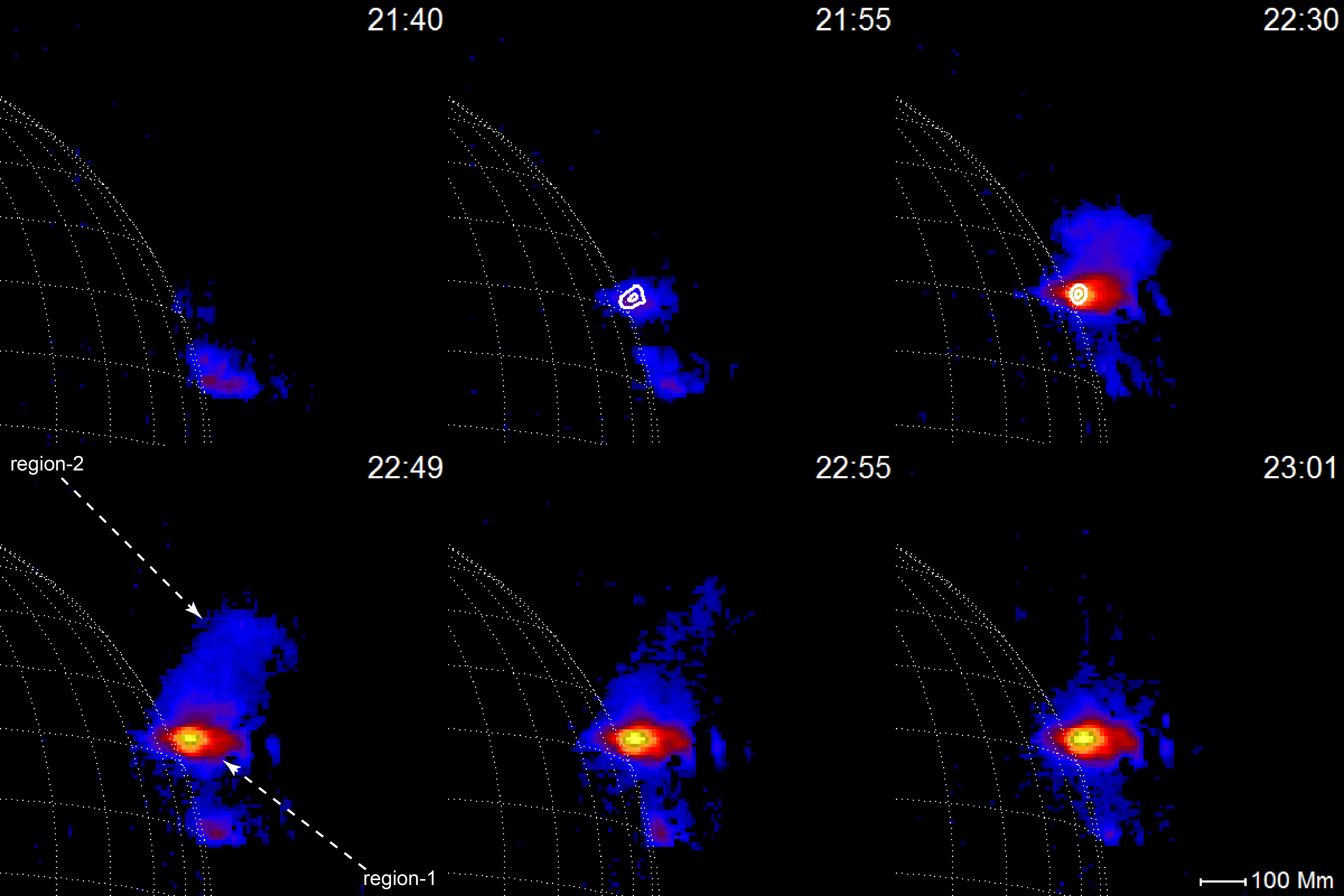}
\caption{Hot plasma dynamics observed with the \ion{Mg}{12} spectroheliograph. Blue corresponds to low intensities, red and yellow correspond to high intensities. Contours mark the location of the RHESSI 6--12~keV emission.}
\label{F:Mg_panel}
\end{figure*}

\begin{figure*}[t]
\centering
\includegraphics[width = \textwidth]{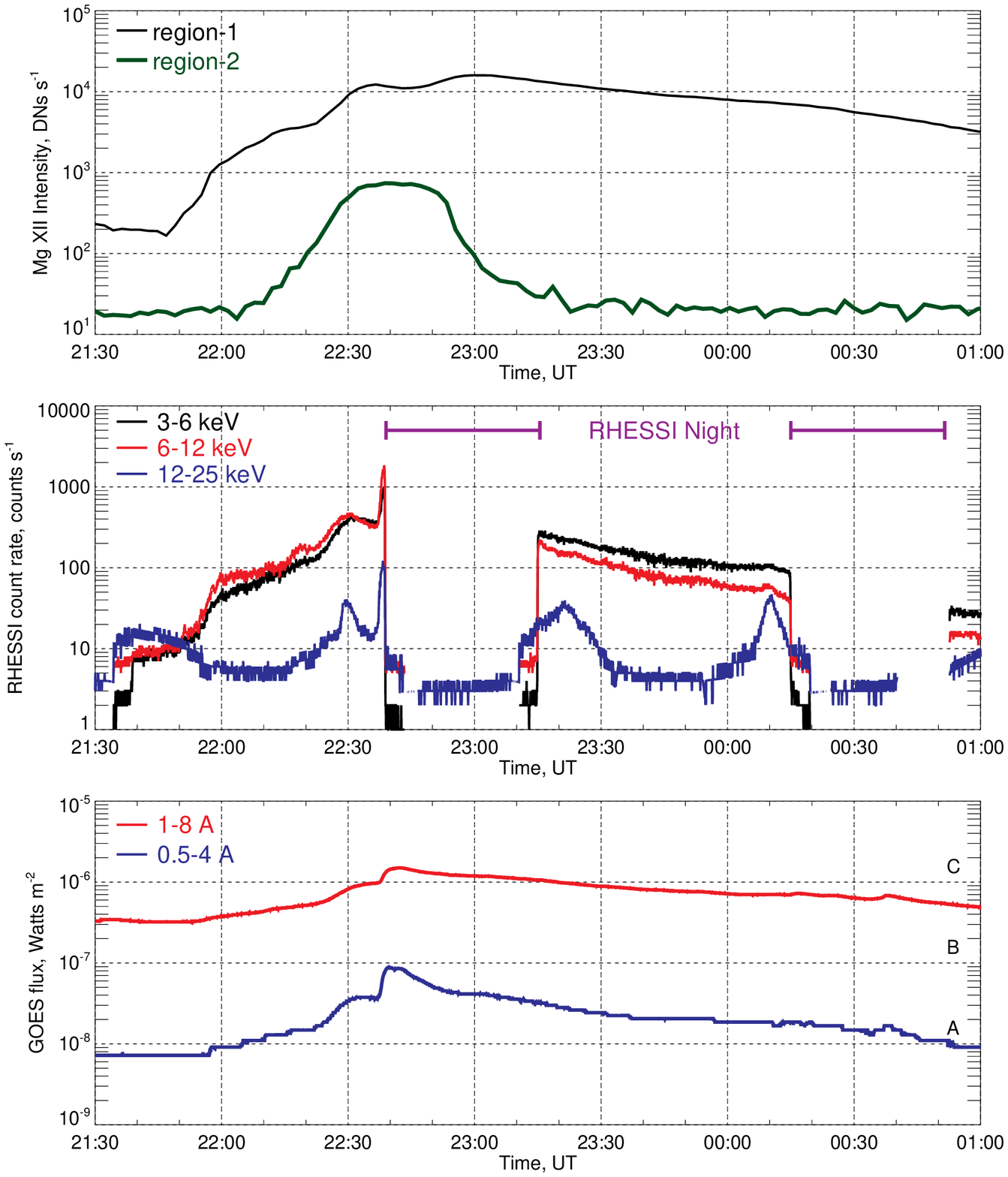}
\caption{X-ray flux dynamics. Top: the \ion{Mg}{12} lightcurves of region-1 (thin black line) and region-2 (thick black line). Middle: the RHESSI count rates. Black line: 3--6~keV channel; red line: 6--12~keV channel; blue line: 12--25~keV channel. Purple lines mark the RHESSI nights. Bottom: GOES flux. Red line: 1--8~\AA\ channel; blue line: 0.5--4~\AA\ channel.}
\label{F:Xray_flux}
\end{figure*}

The observed flare topology is consistent with the standard flare model. We can clearly see a flare arcade, a fluxrope, and a cavity that surrounds the fluxrope (see Figure~\ref{F:trace_eit_mk4}). The CME evolution is also consistent with the standard model. It has a stable phase, an impulsive acceleration phase (probably caused by the impulsive reconnection), and a steady acceleration phase \citep[probably caused by the solar wind;][]{Yashiro2004}. Since the studied flare looks and behaves exactly as the standard model predicts, it is natural to assume that a current sheet should exist between the flare arcade and the CME core.

\subsection{Observation of the Plasma Heating}

As we said above, the main goal of this paper was to find clear evidence of plasma heating in the vicinity of the reconnecting current sheet. For this purpose, we carefully checked all the \ion{Mg}{12} images obtained during the studied event. Thanks to the specific temperature sensitivity of the \ion{Mg}{12} spectroheliograph (starting from $T \ge 4$~MK), we consider the corresponding images as a good marker of plasma heating. As soon as the signal appears in a \ion{Mg}{12} image, we can conclude that the plasma temperature increases to 4~MK or higher. 

The observed dynamics of hot plasma is shown in Figure~\ref{F:Mg_panel}. We found two high-temperature regions during the flare. The first one (the brightest one) appeared at 21:53~UT and existed for $\approx$~5 hours. The region was compact, and its location approximately coincides with the top of the flare loop seen in the TRACE 195~\AA\ image. Plasma heating near the looptop region is a typical feature of a solar flare, and for this reason, we do not consider it in detail. The looptop plasma heating takes place under the current sheet in the so-called `cusp' region of the magnetic configuration. Such a source is often associated with a looptop hard X-ray emission: another typical detail of a solar flare. The HXR emission appears at the same place (see Figure~\ref{F:Mg_panel}) and at the same time (see Figure~\ref{F:Xray_flux}) as region-1.

\begin{figure*}[t]
\centering
\includegraphics[width = \textwidth]{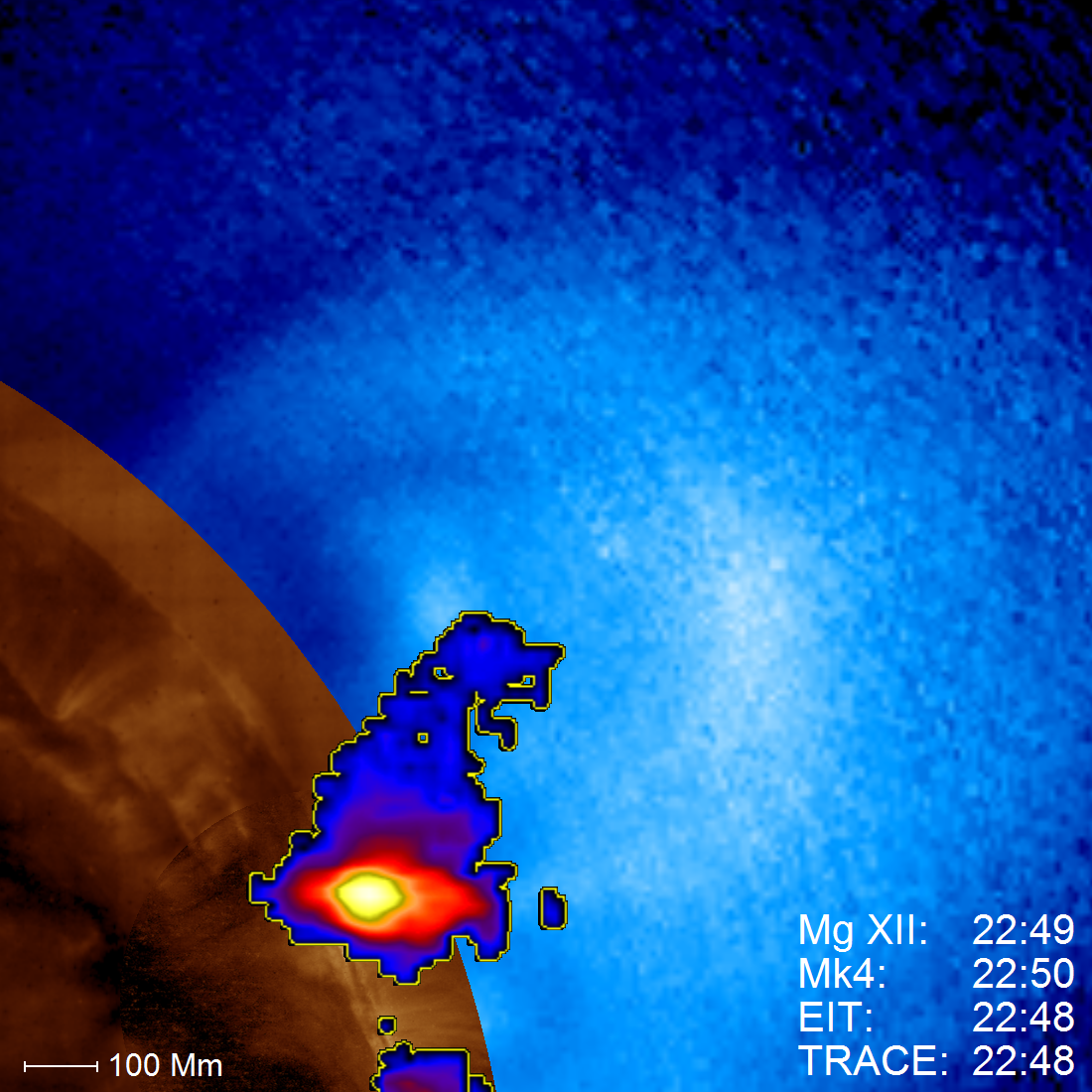}
\caption{Comparison of the \ion{Mg}{12}, TRACE, EIT, and  Mk4 images. Inside yellow contour: signal of the \ion{Mg}{12} spectroheliograph (blue corresponds to low intensities, red and yellow correspond to high intensities). Copper inner circle: TRACE image; copper outer circle: EIT image; blue: Mk4 image. The images were taken on 16 February 2003.}
\label{F:MG_EIT_mk4}
\end{figure*}

The most interesting for us was the second high-temperature region. At 22:28~UT, a faint linear structure appeared above the flare arcade. It gradually increased its length but decreased its intensity. Eventually, the linear structure faded away, reaching the maximal length of 250~Mm. In other parts of the CME, there was no hot plasma. 

\begin{figure*}[t]
\centering
\includegraphics[width = \textwidth]{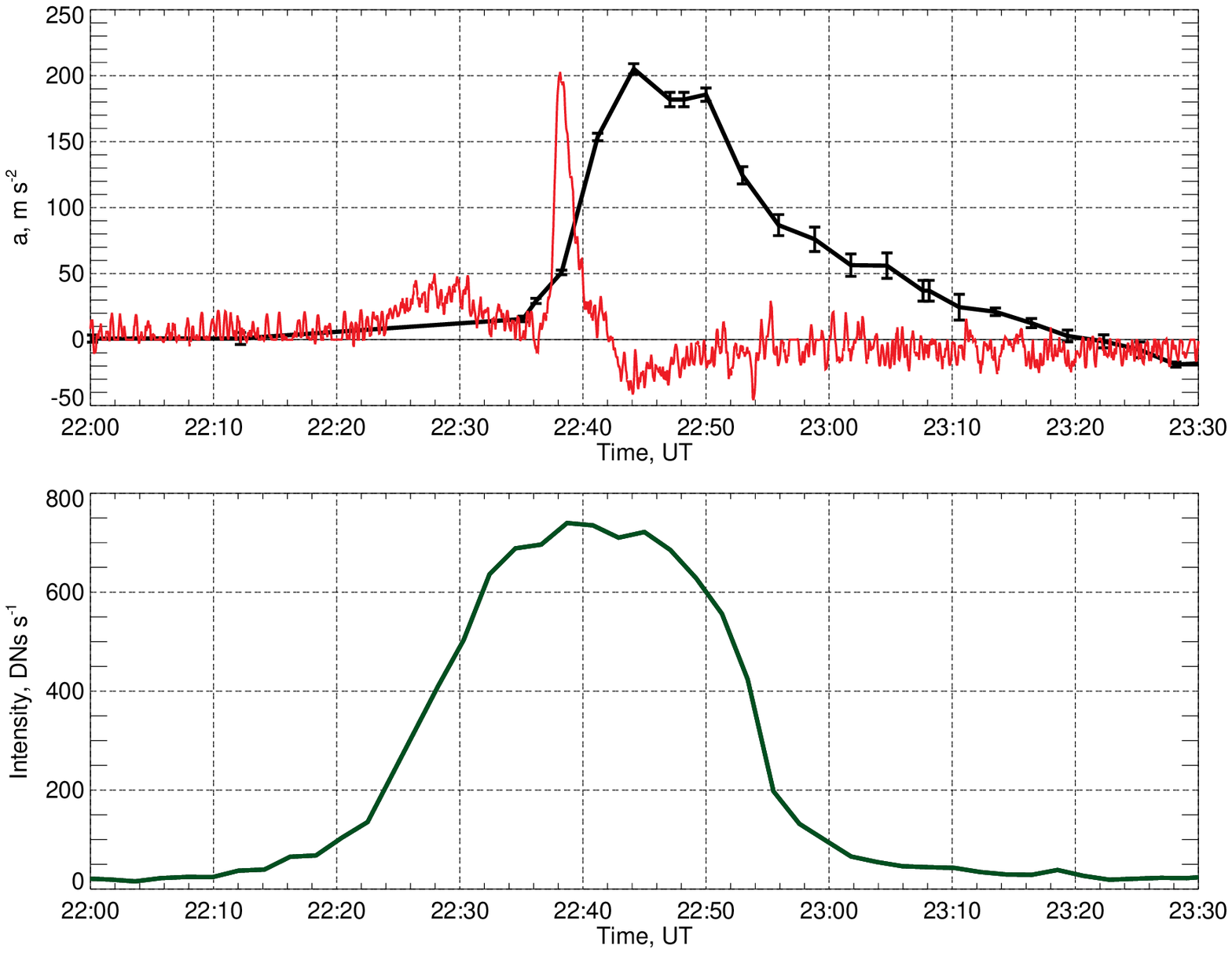}
\caption{Relative timings of the CME's acceleration (black line at the top), the derivative of the GOES 1--8~\AA\ flux (red line at the top), and the linear structure (region-2) intensity in the \ion{Mg}{12} images (green line at the bottom).}
\label{F:relative_timings}
\end{figure*}

The hot linear structure was located between the flare arcade and the CME core (observed in the Mk4 images) and inside the dark cavity observed in the EIT images (see Figure~\ref{F:MG_EIT_mk4}). Such a location coincides with the presumed position of the current sheet that should exist during CME.

Two important features distinguish this second plasma region from the first one. The first feature is the location. Region-2 is clearly not associated with the top of the flare loop and with a `cusp' region. Another feature is a significant difference in brightness and size. Region-2 is much larger than the compact region-1, but, despite this, it is much fainter. The peak flux of the linear structure was $\approx$~4~\% of the peak flux from the compact, bright source (see Figure~\ref{F:Xray_flux}), while the brightness ratio (pixel intensity ratio) was $\approx$~0.5--1~\%. This seriously justifies that region-1 and region-2 were heated by two different mechanisms. We want to emphasize that region-2 is observed simultaneously with region-1. So, we cannot explain them as two consecutive stages of the same heating mechanism.  

The timing of the plasma heating is in good agreement with the CME motion. The flare begins at $\approx$~21:53~UT when the plasma heating starts near the looptop region of the arcade, and the HXR emission starts to rise (see Figure~\ref{F:Xray_flux}). We did not find any signatures of plasma heating in the current sheet (region-2) during this stage. Approximately 30 minutes later, the second stage associated with the CME motion started. At $\approx$~22:30~UT, the CME impulsively accelerated from zero up to $v \approx 300$~km~s$^{-1}$ with an acceleration of $a \approx$~100--200~m~s$^{-2}$. Due to the fast CME motion, the current sheet should be elongating rapidly in the $Z$-direction (see Figure~\ref{F:standard_model}), and just at this time, we observed an additional plasma heating around the presumable location of the current sheet. The CME acceleration lasted about 30 minutes, which is in excellent agreement with the observed duration of plasma heating in region-2 (see Figure~\ref{F:relative_timings}).

Sadly, the RHESSI data were not available during the CME impulsive acceleration (see Figure~\ref{F:Xray_flux}). For this reason, we do not have information about the dynamics of the HXR emission.

\section{Discussion and Conclusion} \label{sec:discussion}

The appearance of high-temperature plasma in the solar corona is typical for solar flares and usually relates to the magnetic reconnection process. Bright sources of thermal X-ray emission usually appear above the top of magnetic loops during the impulsive phase of solar flares. They may be heated by energized electrons or by super-sonic plasma flows.

In a typical magnetic configuration of a flare region, the bright loop-top source is not the only high-temperature region that may appear above the loop during a flare. A much fainter but much larger source may appear higher in the corona due to a diffusion of thermal energy from the region of magnetic reconnection.

Usually, this second source (named region-2 in our study) cannot be observed due to its low brightness, which does not allow it to be distinguished from the low-temperature background. We succeeded in this study due to two main factors. The first one is the region's location and orientation: the CME occurred at the solar limb, and the current sheet was aligned along the line of sight. Such a location and orientation are favorable to detect faint emission of the current sheet. The second one is the temperature sensitivity of the \ion{Mg}{12} spectroheliograph. The instrument detects a high-temperature emission without the contribution of low-temperature background. 

The low brightness ratio of faint and bright components ($\sim$~1~\%) also contributes to the difficulties of the registration. We think that this ratio is typical for flares, but, of course, we have no confirmation of this since only one such event was detected.

The appearance of the faint component we associate with the heating caused by the reconnection inside the current sheet. Comparison of the emission and CME dynamics confirms this idea. The faint high-temperature component appears approximately at the same time as the CME impulsively accelerates. According to the standard CME model, the CME impulsively accelerates during impulsive reconnection inside the current sheet. We think that this clearly indicates that the energy for plasma heating in region-2 comes from magnetic reconnection, which is in good agreement with standard views on solar flares.

Another indirect evidence of the connection between heating and reconnection is the relative timing of region-1 and region-2 emissions. Region-2 appears during the rising phase of the \ion{Mg}{12} and GOES fluxes of region-1 (see Figure~\ref{F:Xray_flux}). According to the Neupert effect \citep{Neupert1968}, the HXR emission of a flare correlates with the derivative of the SXR emission. The GOES 1--8~\AA\ flux derivative---our estimate of the HXR emission---peaks when we observe region-2 (see Figure~\ref{F:relative_timings}). Since HXR emission is a signature of the reconnection, we think that such a correlation further strengthens our interpretation.

The hot plasma was detected in the previous observations of the current sheets. \citet{Ciaravella2002, Ko2003, Ciaravella2008} studied current sheets observed with Ultraviolet Coronagraph Spectrometer \citep[UVCS; ][]{Kohl1995} and reported temperatures around 6~MK. The current sheets analyzed by \citet{Zhu2016} and \citet{Seaton2017} had temperatures around 8--10~MK, while the ones analyzed by \citet{Hannah2013} and \citet{Warren2018} had temperatures around 10--20~MK. Finally, \citet{Landi2012} presented observations, in which current sheet temperature never exceeded 3~MK. Most likely, the difference in the current sheet temperatures is caused by the difference in the reconnection rate.

Our conclusion that the current sheet heating is caused by the reconnection inside it is consistent with the MHD theory and simulations \citep{Yokoyama1998, Seaton2009, Reeves2010}. However, it is difficult to compare this conclusion with other experimental works. Current sheet observations are rare, and only a couple of them allow studying the dynamics of the current sheet heating relative to the dynamics of the reconnection.

The first such example is a current sheet observed during an X8.3 flare on 2017 September 10 \citep{Warren2018}. At the time of writing, this is the most detailed observation of the current sheet \citep[see references in][]{Chen2020}. In this event, some reconnection signatures---the CME impulsive acceleration \citep{Gopalswamy2018, Veronig2018}, the hard X-ray emission, the derivative of the soft X-ray emission, and the microwave emission \citep{Gary2018}---correlated with each other and occurred during a small period of time ($\approx$~15:50--16:00~UT). On the other hand, other reconnection signatures---downflows \citep{Longcope2018} and turbulence \citep{Cheng2018, Warren2018} inside the current sheet---were observed for $\approx$~1~hour after the first reconnection signatures appeared ($\approx$~16:00--17:00~UT). The current sheet itself appeared after the first reconnection signatures ($\approx$~16:00~UT) and hot plasma inside it was observed for $\approx$~1~hour in the 94~\AA\ channel of Solar Ultraviolet Imager \citep[SUVI;][]{Seaton2018}.

Another example is the current sheet observed during the CME on 2009 April 17 \citep{Reva2016b}. The event was observed with the TESIS EUV telescope that build images of the solar corona in the Fe~171~\AA\ line up to distances of 2~$R_\odot$ from the Sun center \citep{Kuzin2011, Reva2014}. The current sheet looked like a double Y-shaped darkening in the Fe~171~\AA\ images. Such darkening indicates that the current sheet is heated (although, it is not clear up to what temperature). The heating (darkening) occurred simultaneously with the CME impulsive acceleration (signature of the reconnection). At the start of the CME impulsive acceleration phase, GOES and SphinX \citep{Gburek2011} registered a short duration flux increase.

All of these three examples---this work, \citet{Reva2016b}, and \citet{Warren2018}---exhibit a similar pattern. During the CME eruption, a short-duration reconnection signatures appear, which are followed by the long-duration reconnection signatures. The heating inside the current sheet starts approximately when the short-duration reconnection signature appears. The hot plasma inside the current sheet is observed during the lifetime of the long-duration reconnection signatures. These similarities support our conclusion that heating inside the current sheet is caused by the reconnection.

However, there are differences between these events. In this work and \citet{Reva2016b}, the heating starts slightly before the short-duration reconnection signatures; in \citet{Warren2018}, it starts after. In this work and \citet{Reva2016b}, the impulsive CME acceleration is a long-duration signature; in \citet{Warren2018}, it is a short-duration signature. These differences and the fact that some reconnection signatures have short duration and some long duration show that the reconnection dynamics has a complex structure, which may vary from one event to another.

We are sure that the energy for the current sheet heating comes from the magnetic reconnection. However, the exact mechanism of the heating is unclear. As we said in the Introduction, ohmic heating \citep{Reeves2010, Reeves2019}, adiabatic compression \citep{Birn2009, Reeves2019}, and turbulence \citep{Ye2020} can heat current sheet. Sadly, our data don't allow us to determine the heating mechanism.

\begin{figure*}[t]
\centering
\includegraphics[width = 0.75\textwidth]{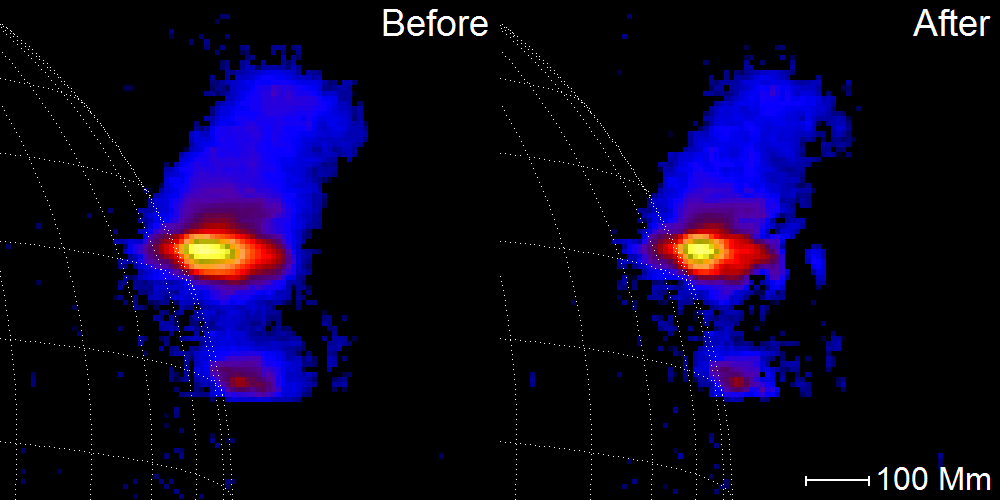}
\caption{Deconvolution of the \ion{Mg}{12} images. Left: image before deconvolution; right: image after deconvolution.}
\label{F:before_after}
\end{figure*}

Another issue that we would like to discuss is the relative weakness of the studied flare. Usually, current sheets are observed during strong flares (X and M classes), while the flare in this work is only C1.4. Figure~\ref{F:halpha_mdi} shows that most of the flaring active region was visible. Even if we take into account partial occultation by the disk, the flare would still be of a C class.

We believe that the studied event is an example of reconnection heating occurring as a universal process in flares from C to X class. Most likely, reconnection heating is less intense in weak flares than in strong ones, and, therefore, plasma is heated to lower temperatures and has lower emission measure. As a result, such heating is rarely observed in weak flares because it is difficult to detect faint hot emission.

We think that monochromatic imagers similar to the \ion{Mg}{12} spectroheliograph---for example, see \citet{Kuznetsov2016, Kirichenko2021, Reva2021}---can help us study reconnection heating. We hope that such instruments will be created in future and that they will improve our understanding of the processes occurring inside the current sheets during solar flares.

\acknowledgments

This is the Accepted Manuscript version of an article accepted for publication in the Astrophysical Journal. IOP Publishing Ltd is not responsible for any errors or omissions in this version of the manuscript or any version derived from it. The Version of Record is available online at doi:10.3847/1538-4357/ac6b3d.

This research was funded by a grant from the Russian Science Foundation (grant No 21-72-10157, https://rscf.ru/project/21-72-10157/).

Mk4 coronameter (DOI: 10.5065/D66972C9) and PICS (DOI: 10.5065/D65719TR) data are provided courtesy of the Mauna Loa Solar Observatory, operated by the High Altitude Observatory, as part of the National Center for Atmospheric Research (NCAR). NCAR is supported by the National Science Foundation. The RHESSI satellite is a NASA Small Explorer (SMEX) mission. SXI full-disk X-ray images are supplied courtesy of the Solar X-ray Imager (SXI) team. The SOHO/LASCO data are produced by a consortium of the Naval Research Laboratory (USA), Max-Planck-Institut fur Aeronomie (Germany), Laboratoire d’Astronomie (France), and the University of Birmingham (UK). MDI and EIT data are supplied courtesy of the SOHO/MDI and SOHO/EIT consortia. SOHO is a project of international cooperation between ESA and NASA.

\newpage

\appendix

\section{\ion{Mg}{12} Deconvolution}

\label{A:prep}

The \ion{Mg}{12} spectroheliograph used Bragg crystal optics. As a result, the instrument has dispersion: its images are a convolution of the spatial component with the profile of the \ion{Mg}{12} 8.42~\AA\ line. More details about this effect could be read in \citet{Kuzin1994} or \citet{Reva2021}.

The \ion{Mg}{12} $\lambda = 8.42$~\AA\ line is a Ly-$\alpha$ doublet of a hydrogen-like Mg ion: $\lambda_1 = 8.4192$~\AA\ ($1s \ {}^2S_{1/2}$ -- $2p \ {}^2P_{3/2}$) and  $\lambda_2 = 8.4246$~\AA\ ($1s \ {}^2S_{1/2}$ -- $2p \ {}^2P_{1/2}$). The ratio of the line intensity should be 2:1.

The \ion{Mg}{12} images consist of two slightly shifted images overlayed one onto another (see Figure~\ref{F:before_after}a). The majority of the pixels in the \ion{Mg}{12} images do not have a signal: they consist of several isolated hot objects. In the direction of the dispersion, the boundaries of those objects contain only a signal from one of the doublet components. The profile of the doublet components overlaps only inside those objects.

To deconvolve the \ion{Mg}{12} images, we use the following algorithm. 
\begin{enumerate}
\item We start at the boundary of the image that corresponds to the short wavelengths.
\item For each pixel, we calculate the intensity of the weaker component (divided by two).
\item We subtract this value from the pixel that corresponds to the location of the weaker component of the doublet.
\item We move in the direction of the dispersion to the next pixel that corresponds to a longer wavelength.
\item We repeat steps 2--4 until we reach the boundary of the image.
\end{enumerate}

The result is shown in Figure~\ref{F:before_after}b. The algorithm successfully eliminates the second component of the doublet. The images become less elongated and easier to interpret.

There are several issues that this algorithm does not address. Firstly, it does not deconvolve the line width. It cannot be deconvolved in a straightforward way because the line width is determined by the Doppler broadening, which varies from pixel to pixel. Secondly, the line ratio of the  Ly$\alpha$ doublet could deviate from the theoretical value \citep{Sylwester1986, lam90}. If this effect is present, it can distort the deconvolved images. Finally, if significant plasma motions along the line of sight are present, the Doppler shifts will distort the images.

\section{CME Kinematics Measurements}

\label{A:kinematics}

For the measurements of the CME coordinates, we used the data of the EIT telescope, Mk4 coronameter, and LASCO coronagraphs. We used a simple point-and-click method. To estimate the errors of the measurements, we repeated the procedure nine times for each image.

In the Mk4 and LASCO images, we aimed at the CME core's center. Since the core was not seen in the EIT images, we aimed for the lowest part of the dark cavity in the EIT images. 

The coordinates measured in the EIT images (lowest part of the dark cavity) and the white-light images (CME's core) are different parts of the CME. We cannot simply combine these coordinates and compute derivatives (velocity and acceleration). Furthermore, the point-and-click method is subjective, and different instruments image corona differently. The center of the core in images obtained by different instruments corresponds to slightly different parts of the CME. In order to compute derivatives, we need to recalculate coordinates measured in the EIT images to the CME's core coordinates and adjust the core coordinates measured by different white-light instruments.

To solve the problem, we adopted the method from \citet{Reva2016}. We assumed that the CME expands proportionally and that the temporal dependence of the heights of different parts of the CME could be linearly scaled. Then we picked scaling coefficients so that temporal dependence of the CME core height seamlessly transitions from one instrument to another.

After scaling the data, we numerically differentiated the radial distance and obtained radial velocity. Then we numerically differentiated velocity and obtained acceleration. For the differentiation, we used the local least-square approximation method \citep{Wood1982, Reva2017}. The result of the measurements is presented in Figure~\ref{F:kinematics}.

\bibliography{mybibl}{}
\bibliographystyle{aasjournal}


\end{document}